\documentclass[]{article}
\usepackage[style=chem-acs,articletitle=true]{biblatex}
\usepackage[parfill]{parskip}

\usepackage{geometry}
\usepackage{float}
\usepackage{siunitx}
\usepackage{etoc}
 \usepackage[font=small, labelfont=bf]{caption}

\newcommand{\beginsupplement}{%
        \setcounter{section}{0}
        \renewcommand{\thesection}{S.\arabic{section}}
        \renewcommand{\thesubsection}{S.\arabic{subsection}}%
        \setcounter{table}{0}
        \renewcommand{\thetable}{S.\arabic{table}}%
        \setcounter{figure}{0}
        \renewcommand{\thefigure}{S.\arabic{figure}}%
        \setcounter{equation}{0}
        \renewcommand{\theequation}{S.\arabic{equation}}%
     }

\usepackage[utf8]{inputenc}
\usepackage{mathtools}
\usepackage{amsmath}
\usepackage{amssymb}
\usepackage{braket}
\usepackage{textcomp}
\usepackage{xcolor}
\usepackage{hyperref}

\title{{\textbf{Filtering and imaging of frequency-degenerate spin waves using nanopositioning of a single-spin sensor}}}
\author{Brecht G. Simon$^{1,\dagger}$, Samer Kurdi$^{1,\dagger}$, Joris J. Carmiggelt$^1$, Michael Borst$^1$, Allard J. Katan$^1$, \\Toeno van der Sar$^{1,*}$}
\date{}

\begin{document}
\maketitle
\begin{flushleft}
\textbf{Affiliations}\\
\small{
$^1$Department of Quantum Nanoscience, Kavli Institute of Nanoscience, Delft University of Technology, 2628 CJ, Delft, The Netherlands\\

\hfill\break
$^\dagger$ These authors contributed equally to this work.\\
$^*$ Corresponding author. Email: t.vandersar@tudelft.nl }
\end{flushleft}

\phantomsection\addcontentsline{toc}{section}{Abstract}
\begin{abstract}
Nitrogen-vacancy (NV) magnetometry is a new technique for imaging spin waves in magnetic materials. It detects spin waves by their microwave magnetic stray fields, which decay evanescently on the scale of the spin-wavelength. Here, we use nanoscale control of a single-NV sensor as a wavelength filter to characterize frequency-degenerate spin waves excited by a microstrip in a thin-film magnetic insulator. With the NV-probe in contact with the magnet, we observe an incoherent mixture of thermal and microwave-driven spin waves. By retracting the tip, we progressively suppress the small-wavelength modes until a single coherent mode emerges from the mixture. In-contact scans at low drive power surprisingly show occupation of the entire iso-frequency contour of the two-dimensional spin-wave dispersion despite our one-dimensional microstrip geometry. Our distance-tunable filter sheds light on the spin-wave band occupation under microwave excitation and opens opportunities for imaging magnon condensates and other coherent spin-wave modes.  
\end{abstract}
\begin{refsection}
\phantomsection\addcontentsline{toc}{section}{Introduction}
Spin waves are collective spin excitations of magnetically ordered materials, with associated quasi-particles called magnons\cite{Rezende2020}. Due to their low damping, spin waves are promising as information carriers in information-technology devices\cite{Chumak2014,Wang2018,Chumak2022,Cornelissen2015}. Techniques to image spin waves aid in studying such devices and realizing their technological potential. As such, several imaging techniques have been developed, with most established techniques based on the spin-dependent scattering of photons\cite{Acremann2000,Sebastian2015,Sluka2019}.

Nitrogen-vacancy (NV) magnetometry images spin waves by their microwave magnetic stray fields. It uses the electronic spin of the NV lattice defect in diamond as a sensor, which can be read out through spin-dependent photoluminescence (PL), is atomic-sized and can stably exist within nanometers from the diamond surface\cite{Rondin2014,Degen2017}. This enables magnetic imaging with nanoscale spatial resolution and high sensitivity. The NV spin allows probing spin-wave spectra with a $\sim$1-MHz frequency resolution through spin lifetime measurements and characterizing spin-wave amplitudes by measuring the NV spin rotation rate\cite{Bertelli2020}. Recently, NV magnetometry has been used to study domain-wall-guided spin-wave modes\cite{Finco2021}, magnon scattering\cite{McCullian2020, Zhou2021,Bertelli2021a}, spin chemical potentials\cite{Du2017}, and frequency combs\cite{Koerner2022}. To enable sensitivity to target spin-wavelengths, accurate control of the NV-sample distance is crucial because the spin-wave stray fields depend exponentially on the distance to the sample at a length scale set by their wavelength.

\begin{figure}[H]
    \centering
    \includegraphics{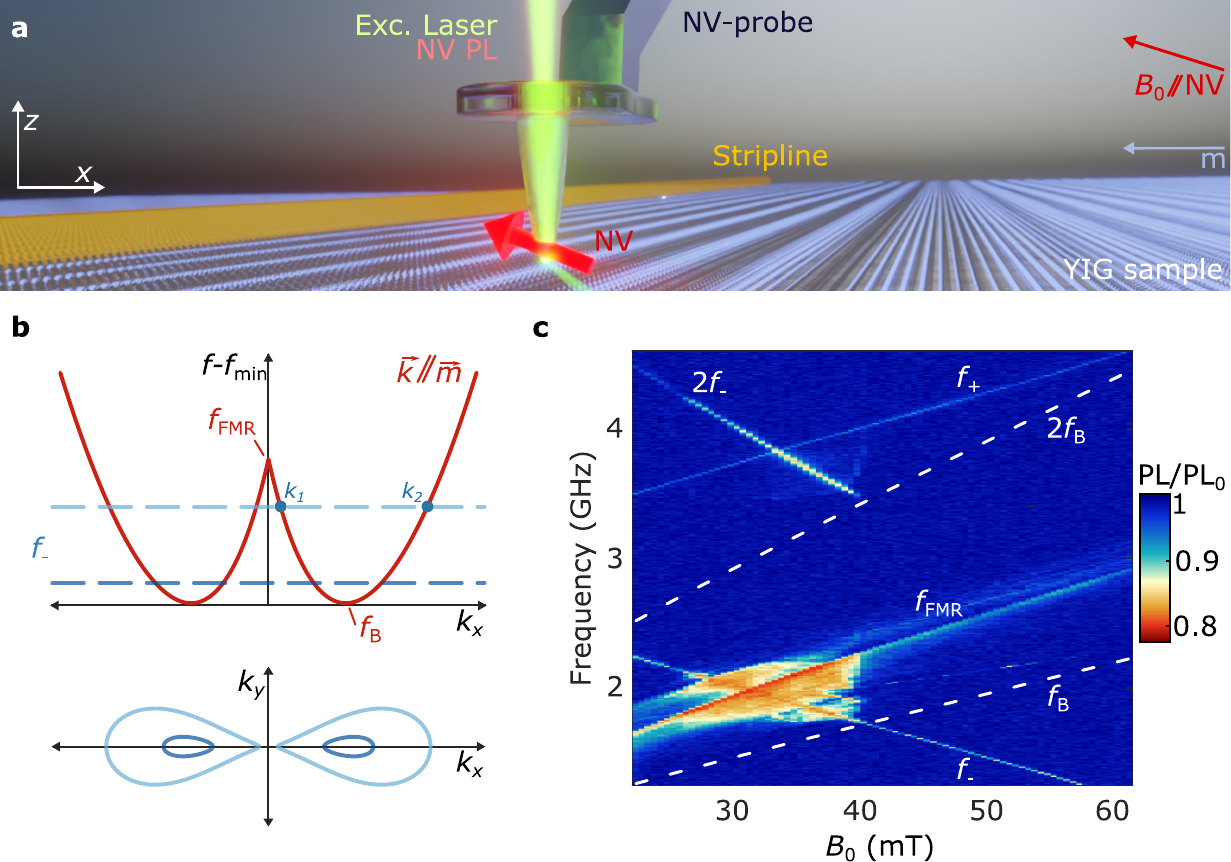}
    \caption{\textbf{Imaging stripline-driven spin waves using scanning nitrogen-vacancy (NV) magnetometry.} (a) A single NV spin embedded $\sim \SI{20}{\nano\m}$ from the apex of a diamond tip measures the magnetic stray fields of spin waves excited by a microwave stripline in a \SI{235}{\nano\m}-thick yttrium iron garnet (YIG) film. The NV spin is initialized using a green laser and read out via its spin-dependent photoluminescence (PL). A bias magnetic field $B_0$ is applied along the NV axis, magnetizing the film perpendicularly to the \SI{1}{\milli\m}-long, \SI{15}{\micro\m}-wide stripline. (b) Top: Calculated dispersion of spin waves traveling parallel to the YIG magnetization (‘backward-volume’ spin waves). The NV spin detects spin waves at its electron spin resonance (ESR) frequency, $f_-$, indicated by dashed lines for two values of $B_0$ (darker color corresponds to a larger $B_0$). In this work we focus on spin waves resonant with $f_-$  in the range $f_\text{B}<f_-<f_\text{FMR}$ for which there exist two frequency-degenerate backward-volume modes, $k_1$ and $k_2$. Bottom: iso-frequency contours of the two-dimensional spin-wave dispersion at the frequencies indicated by the dashed lines in the top panel. (c) NV photoluminescence as a function of $B_0$ and the microwave drive frequency. Data taken with the NV-tip in contact with the YIG at $\sim \SI{30}{\micro\m}$ from the stripline edge at \SI{1}{\milli\watt} drive power. The NV photoluminescence under microwave excitation ($\text{PL}$) is normalized to the NV photoluminescence without microwave excitation ($\text{PL}_0$). The ESR frequencies ($f_{\pm}$) and calculated FMR frequency $f_\text{FMR}$ are labelled. The dashed lines indicate the calculated minimum spin-wave frequency $f_\text{B}$  and its harmonic at $2f_\text{B}$.}
    \label{Fig_1}
\end{figure}

\phantomsection\addcontentsline{toc}{section}{Results}
Here, we demonstrate that controlling the NV-sample distance using a diamond tip mounted on an atomic force microscope (Fig. \ref{Fig_1}a) creates a tunable wavelength filter that enables selective probing of frequency-degenerate spin-wave modes. Increasing the NV-sample distance progressively filters out small-wavelength spin waves, enabling studies of long-wavelength modes that are otherwise hidden in thermal spin-wave noise. We demonstrate high-contrast imaging over a range of wavelengths by adjusting the NV-sample distance on the nanoscale. When maximizing the wavenumber-cutoff of our distance-tunable filter via in-contact scans, we find a surprising pattern of standing spin waves instead of the expected traveling waves. Fourier transforms of the patterns reveal an occupation of spin-wave modes along the entire iso-frequency contour of the two-dimensional spin-wave dispersion despite our one-dimensional stripline geometry, which we attribute to spin-wave scattering. These results show that the exponential decay of the spin-wave stray fields provide a resource unique to magnetic-resonance spin-wave imaging, enabling wavenumber-selective detection of frequency-degenerate spin waves and high-resolution imaging of spin-wave scattering.

Our system consists of a thin film of yttrium iron garnet (YIG), a magnetic insulator with ultra-low spin-wave damping\cite{Serga2010}. We excite spin waves by applying a microwave current to a stripline that is microfabricated onto the YIG surface (Fig. \ref{Fig_1}a, methods). We apply a bias field $B_0$ along the NV axis to tune the NV electron spin resonance (ESR) frequencies ($f_\pm$) relative to the spin-wave band. The orientation of $B_0$ magnetizes the film in-plane and perpendicularly to the stripline, enabling efficient excitation of ‘backward-volume’ spin waves\cite{Serga2010} that travel parallel to the magnetization (Fig. \ref{Fig_1}b) (Supporting Information, Note \ref{sup_sec:Dispersion}-\ref{sup_sec:striplinefield}). The spin waves generate magnetic stray fields above the surface that drive our NV spin when resonant with an NV electron spin resonance (ESR) frequency. We detect these NV-resonant spin waves via the NV-center’s spin-dependent photoluminescence\cite{Wolfe2014} (PL).

We start by providing an overview of the NV PL as a function of the frequency of the microwave current applied to the stripline and the bias field $B_0$ (Fig. \ref{Fig_1}c). We do so with the diamond tip in contact with the YIG at $\sim\SI{30}{\micro\m}$ from the stripline (methods). We observe several regions of reduced PL caused by NV spin transitions that provide a first insight into the spin waves excited by the stripline: First, two lines of reduced PL occur when the drive frequency is resonant with the NV ESR frequencies $f_-$ and $f_+$. Here, the NV spin is driven by the sum of the direct stripline field and the stray field of spin waves excited by the stripline\cite{Bertelli2020, Zhou2021}. Second, a line of reduced PL reveals the YIG ferromagnetic resonance (FMR). Here, FMR-induced magnon-magnon scattering leads to spin-wave noise at the NV frequencies that causes NV spin relaxation and an associated PL reduction\cite{McCullian2020, Du2017}. Third, we observe a broad region of reduced PL when $f_-$ is in the vicinity of the FMR. In this region, the stripline efficiently excites spin waves because of their micron-scale wavelengths near the FMR (Supporting Information, Note \ref{sup_sec:striplinefield}). These spin waves in turn scatter efficiently to modes resonant with $f_-$ because they are close in frequency and wavelength\cite{Hula2020}, causing NV spin relaxation. Correspondingly, the region of reduced PL ends abruptly when $f_-$ drops below the bottom of the spin-wave band (labelled $f_\text{B}$ in Fig. \ref{Fig_1}c) at $B_0\approx 41$ mT. In this work, we study spin waves in the region $f_\text{B}<f_-<f_{\text{FMR}}$ and use the nanoscale control of the NV tip as a wavelength filter to separate the contributions from frequency-degenerate incoherent and coherent spin waves.  

\begin{figure}[H]
    \centering
    \includegraphics{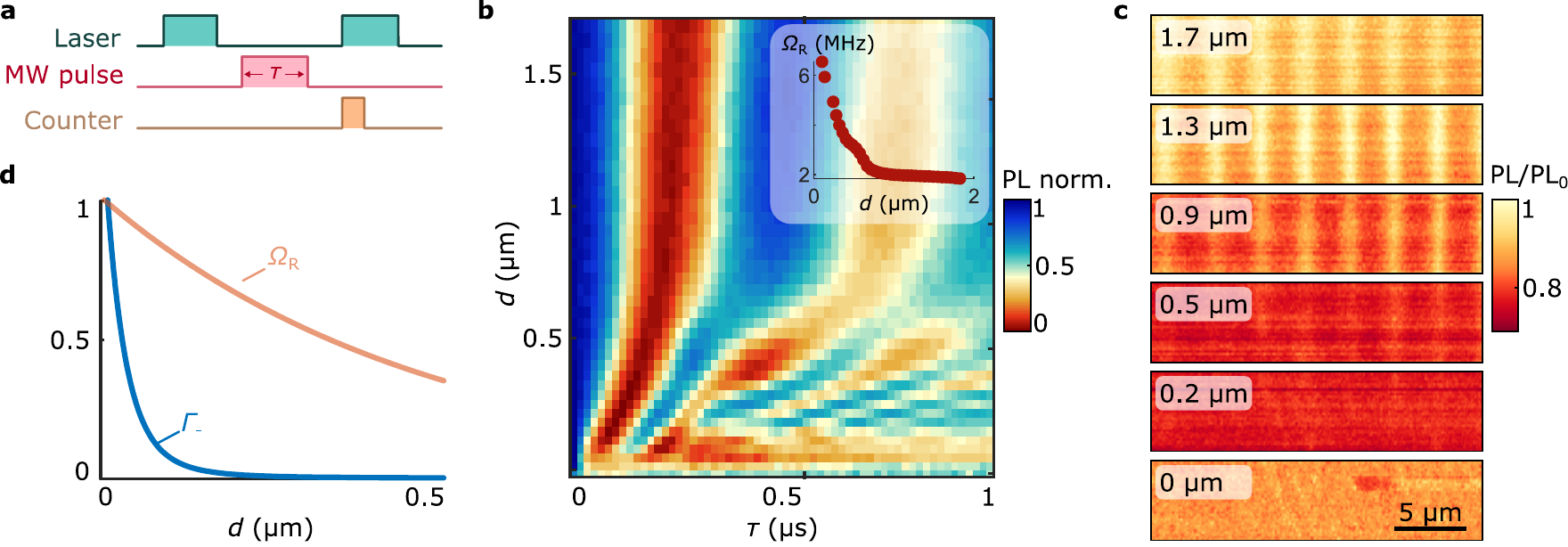}
    \caption{\textbf{Tuning the NV-sample distance as a filter to selectively image a long-wavelength spin-wave mode.} (a) Pulse sequence used for the measurement in (b): A \SI{2.5}{\micro\s} green laser pulse initializes the NV spin. A variable-duration microwave (MW)-pulse excites spin waves. The final NV spin state is read out by measuring the NV photoluminescence during the first \SI{600}{\nano\s} of a second green laser pulse. (b) Spin-wave-driven NV spin dynamics vs tip-sample distance  $d$. The dynamics are governed by the stray-field spectrum of the spin waves at the NV frequency. For $d\gtrsim\SI{100}{\nano\m}$, the stray field of a coherent spin wave yields high-visibility Rabi oscillations with a long decay time. Below $\sim\SI{100}{\nano\m}$, the Rabi decay time starts to vanish, attributed to the more rapidly increasing stray field generated by thermal spin waves (see (d)). Measurement taken at \SI{31}{\micro\m} from the stripline edge at $B_0=\SI{32}{\milli\tesla}$, $f_-=\SI{1.98}{\giga\hertz}$ and $P_\text{MW}= \SI{6.3}{\milli\watt}$. Inset: Fitted Rabi frequency vs $d$ down to \SI{100}{\nano\m}. (c) Spatial maps of the ESR contrast while driving spin waves at $f_-=\SI{1.89}{\giga\hertz}$ and $B_0=\SI{35}{\milli\tesla}$ at  $P_\text{MW} = \SI{1}{\milli\watt}$ for varying $d$. The ESR contrast is obtained by normalizing the NV photoluminescence under microwave excitation ($\text{PL}$) to that without microwave excitation ($\text{PL}_0$). (d) Calculation comparing the NV relaxation rate $\Gamma_-$ caused by thermal spin waves to the NV Rabi oscillation rate caused by a coherently driven spin wave. Both rates are normalized to their value at $d=0$ to highlight the different scaling with distance. The calculation of $\Gamma_-$ assumes an equal population of all spin-wave modes at frequency $f_-$ as expected for a Rayleigh-Jeans distribution\cite{Rustagi2020}. The calculation of $\Omega_\text{R}$ assumes only a single spin-wave mode with wavenumber $k=\SI{2}{\per\micro\m}$ (as imaged in (c)) is excited.}
    \label{Fig_2}
\end{figure}

Spin waves generate a rotating magnetic stray field with amplitude $B_\text{SW}$ that decays with increasing distance $d$ to the sample\cite{Rustagi2020}, with the decay length set by the spin-wavenumber $k$ according to:
\begin{equation}
    B_{\text{SW}} \propto ke^{-kd}
\label{eq:BSW}
\end{equation}
As such, increasing the NV-sample distance progressively filters out the stray fields of high-wavenumber spin waves (Supporting Information, Note \ref{sup_sec:filter}). We demonstrate the filtering by characterizing the stray fields of spin waves excited by the microwave stripline as a function of the NV-sample distance. We do so by measuring the NV spin rotation rate (Rabi frequency), which depends linearly on the amplitude of the NV-resonant microwave field. We measure the Rabi frequency by tuning the NV frequency $f_-$ to the iso-frequency contour of figure \ref{Fig_1}b and applying variable-duration microwave pulses. These pulses excite $f_-$-resonant spin waves that drive NV spin rotations via their magnetic stray field\cite{Andrich2017} (Fig. \ref{Fig_2}a-b). 

With the tip in contact with the YIG (Fig. \ref{Fig_2}b, $d=0$ nm), we observe fast NV spin decoherence, indicating a strong presence of incoherent spin-wave noise. As further shown below, the noise is caused by a combination of thermal and microwave-excited spin wave modes. By lifting the NV a few hundreds of nanometers, we suppress the noise sufficiently and start observing NV Rabi oscillations, indicating a coherent microwave field at the NV frequency. The non-exponential decrease of the Rabi frequency with a further increasing $d$ (Fig. \ref{Fig_2}b and its inset) shows that the Rabi oscillations are driven by an ensemble of coherent spin waves of which the high wavenumbers are progressively suppressed by the distance-dependent cutoff of the filter.

Using spatial maps of the ESR contrast (Fig. \ref{Fig_2}c), we demonstrate that the distance-tunable filter enables spatial imaging of a single low-$k$ spin wave within an ensemble of frequency-degenerate spin-wave modes. We define the contrast $C$ by the ratio of the NV PL with and without microwave drive ($C=1-\text{PL}/\text{PL}_0$). The spatial contrast arises due to the interference between the field of the excited spin waves (which are propagating) and the uniform reference field that is supplied by our stripline\cite{Bertelli2020,Zhou2021}. The in-contact scan (bottom panel Fig. \ref{Fig_2}c) shows two important features: first, the maximum contrast, $C_\text{max} (d=0)=0.15$, is reduced with respect to the maximum contrast at increased distances $C_\text{max} (d>\SI{200}{\nano\m})=0.25$. Second, the contrast equals its maximum value throughout the scan (i.e., it is saturated). We attribute the reduced contrast of the in-contact scan to the strong distance dependence of the stray fields generated by thermally excited spin waves (Fig. \ref{Fig_2}d), which cause NV relaxation and PL reduction in the absence of the microwave drive\cite{Rustagi2020,Flebus2018}(Supporting Information, Note \ref{sup:RateCalc}). The spatially homogeneous saturation indicates a large amplitude of the microwave-driven spin waves, as we will show in more detail below. 

Retracting the tip to $d=\SI{0.2}{\micro\m}$, we find that the contrast approximately doubles with respect to $d=\SI{0}{\micro\m}$. This is expected from the rapid suppression of the thermal spin-wave stray fields by our filter. However, we still find that the contrast is saturated over the entire spatial map (Fig. \ref{Fig_2}c) due to the large stray fields of spin waves excited by the microwave drive. For distances $d>\SI{1}{\micro\m}$, the microwave-driven spin waves are filtered to an extent that yields a clear spatial image of a single low-$k$ spin-wave mode (Fig. \ref{Fig_2}c). These results show how lifting the tip from the surface filters out high-$k$ spin waves, enabling high-contrast imaging of a single low-$k$ spin wave within an ensemble of thermal and coherent spin-wave modes.
\begin{figure}[H]
    \centering
    \includegraphics{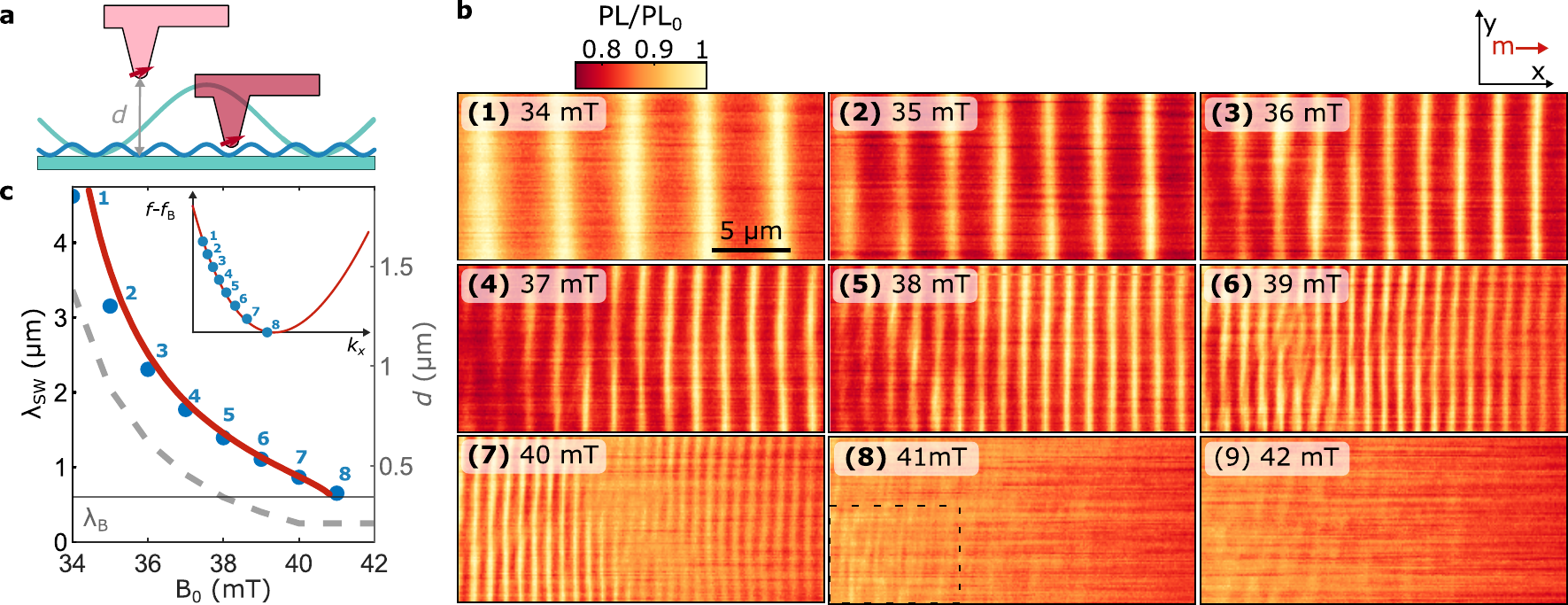}
    \caption{\textbf{Adapting the sensor-to-sample distance to realize high-contrast imaging of different spin-wavelengths.} (a) Spin waves generate a magnetic stray field that decays exponentially at the scale of the spin-wavelength. We tune the tip-sample distance $d$ to optimize the detection of different wavelengths. (b) Spatial maps of the NV ESR contrast showing backward-volume spin waves excited by the stripline at different magnetic fields $B_0$. Increasing $B_0$ (panels 1-8) decreases the wavelength of the spin waves that are resonant with the NV ESR frequency. In each scan, we tune the distance $d$ to maintain a constant ESR contrast, as plotted in c. In panel 8, we extract the wavelength by analyzing the dashed box. Drive power $P_\text{MW}= \SI{4}{\milli\watt}$. (c) The wavelengths extracted by fitting (Supporting Information Note \ref{sup_sec:spatial_Cesr}) the wave patterns in b (blue dots) compared to the wavelengths calculated from the backward-volume dispersion (plotted in red, Supporting Information, Note \ref{sup_sec:Dispersion}), as a function of $B_0$. The wavelength at the minimum of the spin-wave band is indicated by $\lambda_\text{B}$. Grey dashed line (right y-axis): The tip-sample distance $d$ used in each of the scans shown in b. Inset, red line: calculated dispersion of the backward-volume spin waves relative to the minimum spin-wave frequency. Blue dots: modes imaged in b.}
    \label{Fig_3}
\end{figure}
Fast NV-imaging of spin waves requires a strong ESR contrast. Because the spin-wave stray field falls off exponentially (Eq. \ref{eq:BSW}), maintaining a strong contrast requires adapting the NV-sample distance to the expected spin-wavelength (Fig. \ref{Fig_3}a). We change the wavelength of the mode, indicated by $k_1$ in figure \ref{Fig_1}b, by increasing $B_0$ while reducing the drive frequency according to $f_- = D-\gamma B_0$ to maintain resonance with the NV, where $D=\SI{2.87}{\giga\hertz}$ is the NV zero-field splitting and $\gamma=\SI{28}{\giga\hertz\per\tesla}$ is the electron gyromagnetic ratio. Starting from the distance used in figure \ref{Fig_2}c for $B_0 = \SI{35}{\milli\tesla}$, we find that keeping $kd=constant$ yields high-contrast images over a range of wavelengths (Fig. \ref{Fig_3}b). The spatial images of figure \ref{Fig_3}b show how the wavelength decreases with increasing $B_0$ until the $f_-$ detection frequency drops below the bottom of the spin-wave band at $B_0\approx \SI{41}{\milli\tesla}$ (inset Fig. \ref{Fig_3}c). The large ESR contrast enables a straightforward extraction of the wavelengths (Supporting Information Note \ref{sup_sec:spatial_Cesr}), which correspond well with the calculated spin-wave dispersion (Fig. \ref{Fig_3}c).

\begin{figure}[H]
    \centering
    \includegraphics{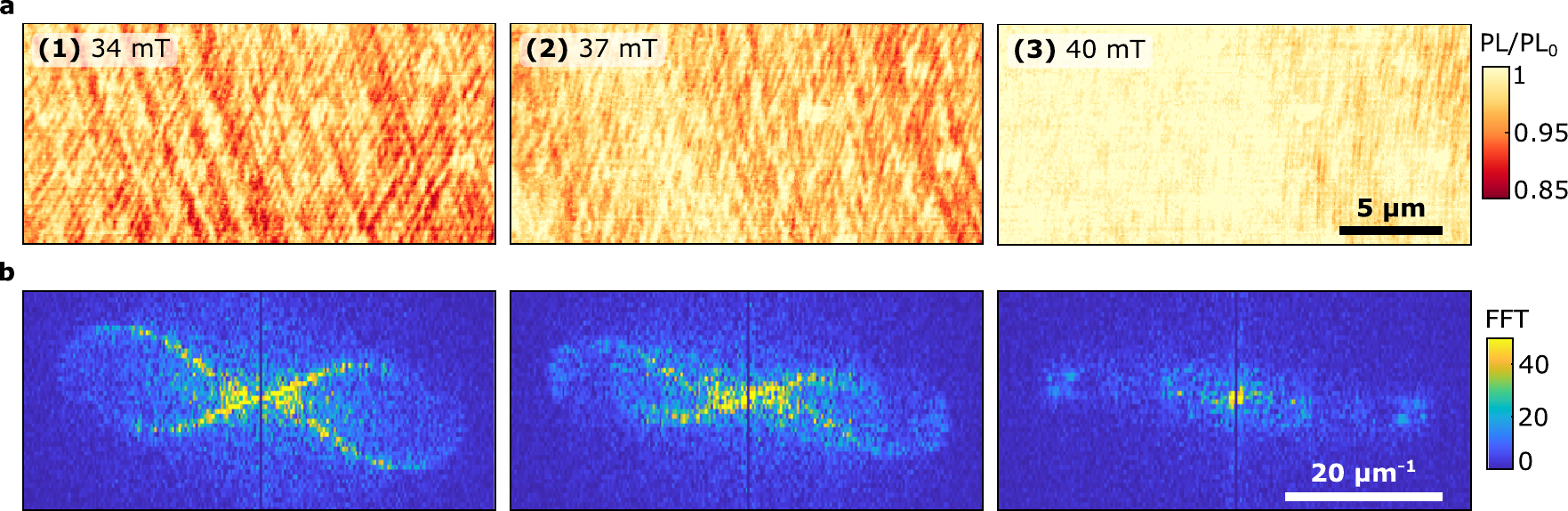}
    \caption{\textbf{Emergence of frequency-degenerate, standing spin-wave modes in spatial maps of the ESR contrast.} (a) Measured ESR contrast when the tip is in contact with the sample at low drive power (\SI{6.3}{\micro\watt}) for different magnetic bias fields $B_0$. (b) Absolute value of the Fourier-transformations of the maps in (a), revealing the wavevectors present in the spatial spin-wave patterns. The full isofrequency contour of the spin-wave dispersion is visible at $B_0=\SI{34}{\milli\tesla}$.}
    \label{Fig_4}
\end{figure}

Bringing the NV-tip into contact with the sample maximizes the wavenumber cutoff of our filter and increases the relative contribution of high-wavenumber modes to the stray field (Supporting Information, Note \ref{sup_sec:filter}). We use in-contact scans to study the ensemble of spin-wave modes excited in the magnetic film. To avoid the in-contact saturation observed in figure \ref{Fig_2}c, we tune down the drive power by a factor 500, which reveals a rich pattern of spin waves in different directions (Fig. \ref{Fig_4}a). 

To interpret the wavenumber content of the spin-wave patterns observed in figure \ref{Fig_4}a, we perform a Fourier transform. The Fourier maps reveal the excitation of spin-wave modes along the entire $f_-$-isofrequency contour of the two-dimensional spin-wave dispersion (Fig. \ref{Fig_4}b). Although these modes are not directly excited by our microstrip, such an homogeneous occupation of the spin-wave dispersion may be expected when taking into account scattering of the primarily excited backward-volume spin waves\cite{Mohseni2019} enhanced by the presence of defects in our film\cite{Gross2020,Graefe2020} (Supporting Information, Note \ref{sup_sec:yig_surface}). 

Surprisingly, the absence of ESR contrast for fields larger than \SI{40}{\milli\tesla} (Supporting Information, Note \ref{sup_sec:overview}) shows that the amplitude of the direct stripline field is insufficient to generate the observed standing-wave pattern in the magnetic stray field. We therefore conclude that the observed spin waves are standing waves, created by scattered waves that have a fixed phase relation with the stripline drive field. These results highlight the coherent nature of the scattering process and the efficiency by which it leads to the occupation of high-momentum modes that are otherwise inaccessible to a one-dimensional excitation stripline. 

\phantomsection\addcontentsline{toc}{section}{Discussion/Conclusion}
Nanoscale control of the NV-sample distance serves as a tunable filter that enables balancing the magnetic fields generated by an ensemble of incoherent and coherently driven spin waves of different wavelengths. This control enables selective imaging of a coherent spin-wave mode within a mixture of frequency-degenerate spin waves and retaining a high-visibility response when imaging different wavelengths. In-contact scans at reduced drive power show a surprising pattern of standing spin-wave modes. The Fourier transforms of these patterns reveal spin-wave occupation along the entire iso-frequency contour of the two-dimensional spin-wave dispersion. We attribute the occupation of these high-momentum modes to defect-enhanced spin-wave scattering. The phase relation between the scattered modes is maintained, emphasizing the coherent nature of the scattering process. Nanoscale control of the NV-sample distance and wavenumber-selective imaging of magnetic oscillations at microwave frequencies paves the way for imaging magnon condensates\cite{Demokritov2006} or other coherent spin-wave modes\cite{Finco2021}, and could also be used to probe microwave electric current distributions in devices.

\subsection*{Methods}\addcontentsline{toc}{section}{Methods}
\subsubsection*{YIG Sample}
The $\sim\SI{235 \pm 10}{\nano\m}$ thick yttrium iron garnet (YIG) was grown on a gadolinium gallium garnet (GGG) substrate by liquid-phase epitaxy (Matesy GmbH). The YIG chip was first sonicated in acetone to remove contaminants. A \SI{1}{\milli\m}-long and \SI{15}{\micro\m}-wide stripline (\SI{5}{\nano\m} titanium / \SI{200}{\nano\m} gold) for spin-wave excitation was then deposited on top of the YIG surface using \textit{e}-beam evaporation preceded by \textit{e}-beam lithography, using a double PMMA resist (A8 495K / A3 950K) and a top layer of Elektra95.

\subsubsection*{Measurement Setup}
Our scanning NV-magnetometry setup consists of two stacks of Attocube positioners (ANPx51/RES/LT) and scanners (ANSxy50/LT and ANSz50/LT) that enable individual positioning of the tip and sample, in addition to a confocal microscope setup, which are all placed in an acoustical enclosure. The confocal setup uses a \SI{515}{\nano\m} green laser (Cobolt 06-MLD, pigtailed) for NV excitation, which is focused by the objective lens (LT-APO/VISIR/0.82) onto a single-NV tip with the NV located approximately \SI{20}{\nano\m} below the tip surface ((001)-oriented, Qzabre). The NV photoluminescence (PL) is collected by the same objective and separated from the excitation laser by a dichroic mirror (Semrock Di03-R532-t3-25x36) and a long-pass filter (Semrock BLP01-594R-25), spatially filtered by a pinhole (\SI{50}{\micro\m}), and finally collected by an avalanche photodiode (APD) (Excelitas SPCM-AQRH-13). A SynthHD (v2) microwave generator (Windfreak Technologies, LLC) was used to apply microwave signals. A programmable pulse generator (SpinCore Technologies, Inc. PulseBlasterESR-PRO 500) controls the timing of the laser excitation, detection window and microwaves. A National Instruments card (PCIe 6323) was used for the data acquisition. 

\subsubsection*{Spin-wave imaging}
All measurements were performed close to the middle of the \SI{1}{\milli\m}-long stripline to prevent edge effects from stripline corners, within \SI{30}{\micro\m} from the edge of the \SI{15}{\micro\m}-wide stripline. The direct stripline field interferes with the spin-wave field to form the standing-wave stray-field patterns of figures \ref{Fig_2} and \ref{Fig_3} \cite{Bertelli2020, Zhou2021}. The static field $B_0$ is applied by moving a small permanent magnet mounted on translation stages. For all measurements, the magnet is aligned along the NV axis within $\sim\SI{5}{\degree}$ such that the expected angle between the NV and the sample is $\theta\approx\SI{54}{\degree}$ with respect to the sample-plane normal. Because of some uncertainty introduced when mounting the NV-probe, we leave this angle as a free parameter when fitting the measured wavelength to the spin-wave dispersion, yielding $\theta=\SI{49}{\degree}$ (Fig. \ref{Fig_3}d). For the scans at non-zero tip-sample distances in figures \ref{Fig_2} and \ref{Fig_3}, we first touch down onto the sample with the tip to acquire a well-defined distance reference. Then, we turn off the AFM feedback and set the lift height using our piezo scanner (Supporting Information, Note \ref{sup_sec:calibration}). We repeat this for each line trace. As there is no feedback, the lift height can change over a line trace due to drift or sample tilt. 

\subsection*{Acknowledgements}\addcontentsline{toc}{section}{Acknowledgements}
The authors thank Yaroslav Blanter for useful discussions.\\
\textbf{Funding:} This work was supported by the Dutch Research Council (NWO) through the NWO Projectruimte grant 680.91.115 and the Kavli Institute of Nanoscience Delft. \\
\textbf{Author contributions:} B.G.S, S.K., A.K. and T.v.d.S. conceived and designed the experiments. B.G.S, S.K., M.B., A.K.  realized the imaging setup. B.G.S., S.K., and A.K performed the experiments. B.G.S., S.K., J.J.C, T.v.d.S. analyzed and modelled the results. S.K. fabricated the stripline on the YIG sample. B.G.S., S.K., and T.v.d.S wrote the manuscript with contributions from all coauthors. \\
\textbf{Competing interests:} The authors declare that they have no competing interests. \\
\textbf{Data availability:} All data contained in the figures will be made available at zenodo.org upon publication with the identifier 10.5281/zenodo.6703953. Additional data related to this paper may be requested from the authors.

\phantomsection\addcontentsline{toc}{section}{References}
\printbibliography
\end{refsection}

\newpage
\beginsupplement 
\begin{refsection} 
\textbf{{\huge Supporting Information}}\hfill\break
\phantomsection
\addcontentsline{toc}{section}{Supporting Information}
\localtableofcontents

\subsection{Spin-wave dispersion}\label{sup_sec:Dispersion}
\begin{figure}[H]
    \centering
    \includegraphics{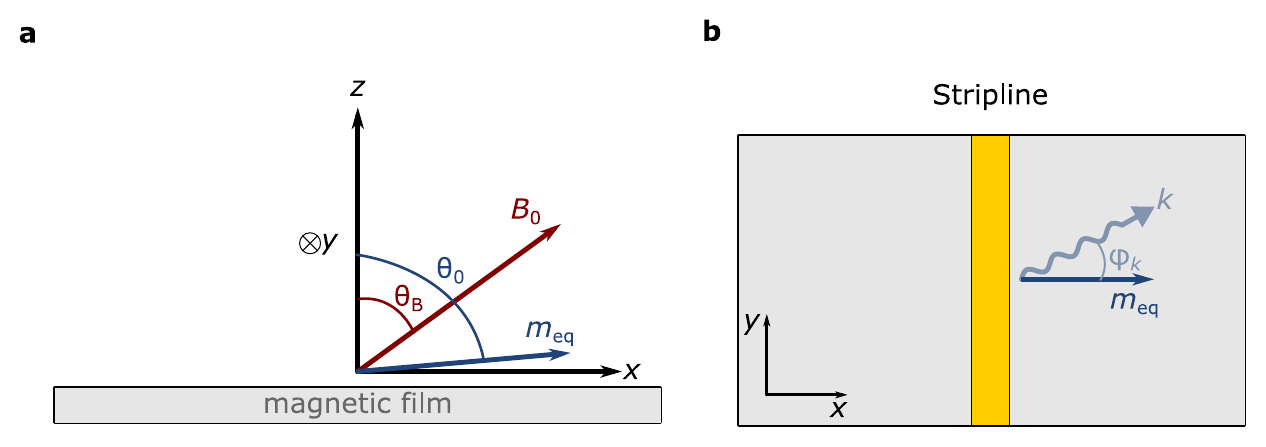}
    \caption{\textbf{Schematic of the measurement geometry.} (a) Side view of our measurement geometry. The magnetic field is applied at an angle $\theta_B$ with respect to the sample normal. As a result the equilbrium magnetization $\vec{m}_{\mathrm{eq}}$ tilts slightly out-of-plane with an angle $\theta_0$. (b) Top view of the measurement geometry. We drive the transverse magnetization via an oscillating magnetic field supplied by a microwave current that is sent through a microstrip with width \SI{15}{\micro\m} and length \SI{1}{\milli\m}. In this work, the stripline field excites spin waves that travel parallel to the equilibrium magnetization, also called backward-volume spin waves.The parameters used for calculating the spin-wave dispersion for the film studied in this work are: $M_s = 1.42\cdot 10^5$ A/m, $A_{\mathrm{ex}}=3.8\cdot 10^{-12}$ J/m, $\alpha=1\cdot 10^{-4}$ and $L = 235$ nm \cite{Bertelli2020}. The angle between the magnetic field and the film, $\theta_B$ is $\theta_B\approx\theta_{\mathrm{NV}}\approx 54$\textdegree{}.}
    \label{sup_fig:geometry}
\end{figure}

Here, we calculate the spin-wave dispersion of our 235 nm film of yttrium iron garnet (YIG). We assume a 2D geometry, where the magnetization does not change across the film thickness (Fig. \ref{sup_fig:geometry}). We first consider the relevant energy contributions for our magnetic system to evaluate the Landau-Liftshitz-Gilbert (LLG) equation that describes the dynamics of the magnetization. Following the approach described by Rustagi \textit{et. al}\cite{Rustagi2020}, we then obtain the magnetic susceptibility (section \ref{sup_sec:susceptibility}) and the spin-wave dispersion (section \ref{sup_sec:dispersion}). 

\subsubsection{Magnetic susceptibility}\label{sup_sec:susceptibility}
Given that the Zeeman interaction, the demagnetizing field and the exchange interaction are the relevant energy contributions, we calculate the response of the transverse magnetization $\delta\vec{m'}_\perp$ to a drive field $\vec{h}_\perp(\vec{k})$ via $\delta\vec{m'}_\perp = S\vec{h}_\perp$. Here, $\delta\vec{m}'$ is defined in the magnet frame, where the equilibrium magnetization ($\vec{m}_{\mathrm{eq}}$) points in the $z$-direction. And $S$ is the transverse magnetic susceptibility, which is given by \cite{Rustagi2020}: 

\begin{align}
    S (\vec{k}, \omega)
        =
        \frac{\gamma}{\Lambda}
        \begin{bmatrix}
            \omega_3 - i\alpha \omega & -\omega_1 - i\omega \\
            -\omega_1 + i\omega       & \omega_2 - i\alpha \omega 
        \end{bmatrix}
    \label{sup_eq:susceptibility_matrix}
\end{align}
where 
\begin{align}\label{eq:omegas}
    \omega_0 (\vec{k})
        &=
        \omega_\text{B} \cos(\theta_B- \theta_0) - \omega_\text{M} \cos^2 \theta_0 + \omega_D k^2, \\
    \omega_1 (\vec{k})
        &=
        \omega_\text{M} f_L \sin \phi_k \cos \phi_k \cos \theta_0, \\
    \omega_2 (\vec{k})
        &=
        \omega_0 + \omega_\text{M} \left[ f_L \cos^2 \phi_k \cos^2 \theta_0 + (1- f_L) \sin^2 \theta_0 \right], \\
    \omega_3 (\vec{k})
        &=
        \omega_0 + \omega_\text{M} f_L \sin^2 \phi_k, \\
    \Lambda (\omega)
        &=
        (\omega_2 - i\alpha \omega)(\omega_3 - i\alpha \omega) - \omega_1^2 - \omega^2.
\end{align}
Here, $\omega_B = \gamma B_0$, is the frequency associated with the Zeeman energy, where $\gamma$ is the gyromagnetic ratio and $B_0$ the externally applied magnetic field. We apply $B_0$ at an angle $\theta_B$, which is the direction of the magnetic field with respect to the sample normal, such that it aligns with the NV center. As a result, the equilibrium magnetization $\theta_0$ tilts out-of-plane by an angle $\theta_0$. 
Next, the frequency associated with the demagnetizing field is given by: $\omega_M = \gamma \mu_0M_s$, where $\mu_0$ and $M_s$ are the vacuum permeability and the saturation magnetization respectively. Finally, $\omega_D=\frac{\gamma D}{M_s}$ is associated with the exchange interaction, where $D$ is the spin stiffness\footnote{The spin stiffness is often expressed in terms of the exchange constant $A$ with more conventional units (J/m): $D = 2\gamma A_{\mathrm{ex}}/M_s$ (with units (rad/s/m$^2$))}. A wave vector, $\vec{k}$ is described by its wavenumber $k$ (i.e. the modulus of the wave vector) and by its direction which is described by $\phi_k$. Finally, the prefactor $f_L$ is given by $f_L \equiv 1-(1-e^{-kL})/(kL)$ in which $L$ is the film thickness.

\subsubsection{Equilibrium magnetization}
The equilibrium angle of the magnetization, $\theta_0$, follows from minimizing the free energy and solving for each value of the magnetic field: 
\begin{equation}
    -2B_0\sin(\theta_B - \theta_0)=\mu_0M_s\sin(2\theta_0)
\label{sup:theta0}
\end{equation}
We calculate that $\theta_0$ is in-plane to within a few degrees for the magnetic fields used in our measurements.

\subsubsection{Spin-wave dispersion}\label{sup_sec:dispersion}
\begin{figure}[H]
    \centering
    \includegraphics{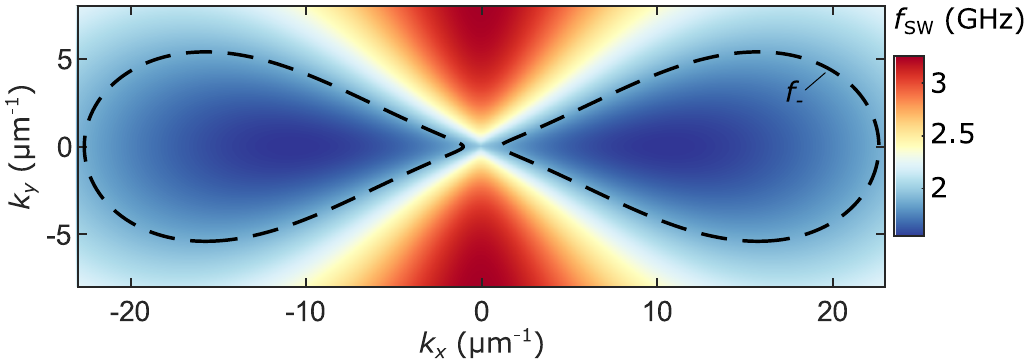}
    \caption{\textbf{Spin-wave dispersion.} Calculated spin-wave dispersion for $B_0=\SI{35}{\milli\tesla}$, i.e. when $f_{\mathrm{B}} < f_- < f_{\mathrm{FMR}}$. The field is aligned along the NV-axis such that $f_- = 1.89$ \si{\giga\hertz}, which is represented by the iso-frequency contour (dashed line).}
    \label{sup_fig:dispersion}
\end{figure}

The spin-wave dispersion is given by the frequencies for which the susceptibility is singular, i.e. when: $\Lambda = 0$ (Fig. \ref{sup_fig:dispersion}).By tuning the external magnetic field, we vary $f_-$ with respect to the minimum spin-wave frequency, as such the contour (dashed line in Fig. \ref{sup_fig:dispersion}) changes shape such that $f_-$ becomes resonant with spin waves of different wavevectors.

\subsection{Stripline field}\label{sup_sec:striplinefield}
\begin{figure}[H]
    \centering
    \includegraphics{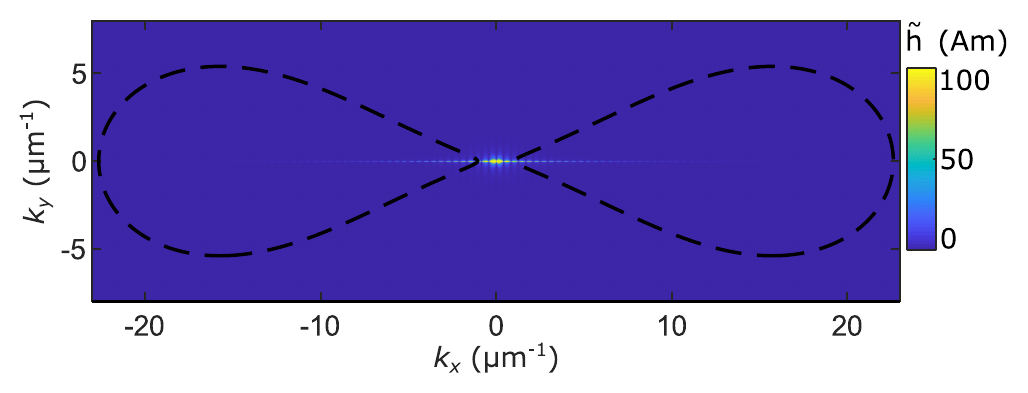}
    \caption{\textbf{The effective field strength for a stripline that is aligned perpendicular to the magnetization (backward-volume geometry).} Using a \SI{15}{\micro\m}-wide and \SI{1}{\milli\m}-long stripline. The stripline field is most efficient in driving low-wavenumber modes (close to the FMR). Due to its one-dimensional character, spin waves with a component in the $k_y$ direction are not excited. Dashed line indicates the $f_-$-isofrequency contour (Fig. \ref{sup_fig:dispersion}) of NV resonant modes at 35 mT.}
    \label{sup_fig:striplinefield}
\end{figure}

We use a stripline oriented along $y$, with width $w$, length $L$ and thickness $h$ for spin-wave excitation, centered at $x=0$ and $z=-h/2$. A microwave current density $J(\omega)$ applied to the stripline generates a magnetic field with components \cite{Bertelli2020}:  
\begin{align}
    h_{\mathrm{x}} = 2J(\omega)e^{kz}\frac{e^{kh}-1}{kk_{x}}\sin(k_{x}\frac{w}{2})\frac{\sin(k_{y} \frac{L}{2})}{k_{y(x)}}\\
    h_{\mathrm{z}} = -2iJ(\omega)e^{kz}\frac{e^{kh}-1}{k^2}\sin(k_{x}\frac{w}{2})\frac{\sin(k_{y} \frac{L}{2})}{k_{y}}\\ \label{sup_eq:hz}
\end{align}
in k-space. Because the film is magnetized along $x$, the x-component of the field does not contribute to spin-wave excitation. As such, we only consider the $z-$component. The field exciting the spin waves is obtained by averaging over the film thickness:
\begin{align}
    \tilde{h}_{z} &= -2iJ(\omega)\frac{e^{-kL}-1}{kL}\frac{e^{kh}-1}{k^2}\sin(k_x\frac{w}{2})\frac{\sin(k_y \frac{L}{2})}{k_y} \label{sup_eq:hz_integr}
\end{align}
Because the length of the stripline far exceeds its width and the distance between the stripline center and our measurement location, it is essentially a one-dimensional stripline that does not excite spin waves in the $k_y$ direction.  (Fig. \ref{sup_fig:striplinefield}).


\subsection{Wavenumber-dependent filtering of the spin-wave stray field}\label{sup_sec:filter}
\begin{figure}[H]
    \centering
    \includegraphics{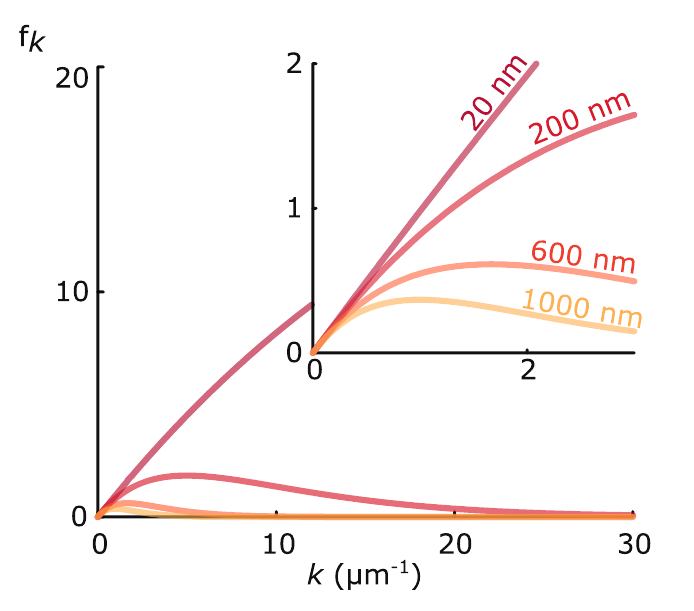}
    \caption{\textbf{Wavenumber selective filtering by tuning the NV-to-sample distance $d$}. The filter function $f_k$ peaks at $k=1/d$, where $k$ is the spin-wavenumber. The filter function is plotted for several values of the NV-to-sample distance}
    \label{sup_fig:SW_filtering}
\end{figure}

The measured spin-wave stray field is proportional to a prefactor $f_k$ that is dependent on the NV-to-sample distance ($d$) according to:
\begin{equation}
    B_{\mathrm{SW}} \propto f_k =  k\exp(-kd)
\end{equation}
Here, $f_k$ is the 'filter function' responsible for the wavenumber-filtering action of the measured spin-wave stray fields. The filter peaks at $k=1/d$ (Fig. \ref{sup_fig:SW_filtering}). We see that increasing the NV-sample distance progressively filters out the stray fields of the short wavelength spin-wave modes.


\subsection{NV relaxation induced by thermal magnons}\label{sup:RateCalc}
We follow the approach of Rustagi \textit{et al.}\cite{Rustagi2020} to calculate the NV relaxation rates induced by the magnons in our YIG film (Fig. \ref{Fig_2}d, main text) using:
\begin{align}
    \Gamma_\mp (\omega_\mp) 
        = 
        \frac{\gamma^2}{2} \int \frac{d\vec{k}}{(2\pi)^2} \sum_{{i,j} \in \{x,y\}} \mathcal{D}^\text{eff}_{\pm i}(\vec{k}) \mathcal{D}^\text{eff}_{\mp j}(-\vec{k}) C_{ij} (\vec{k}, \omega_\mp).
    \label{eq:NV_relaxation_rate}
\end{align}
Here, $\Gamma_{\mp}$ are the relaxation rates corresponding to the $\omega_{\mp}$ ESR  frequencies, $\vec{k}$ is the spin-wavevector,  $\mathcal{C}$ is a spin-spin correlator describing the thermal magnon fluctuations, and $\mathcal{D}^\text{eff}$ is a dipolar tensor that calculates the magnetic stray fields that induce NV spin relaxation generated by these fluctuations. Note, this equation is defined in the magnet frame, for which the equilibrium magnetization is along the $z$-direction.

The thermal transverse spin fluctuations in the film are described by \cite{Rustagi2020}: 
\begin{align}
    C_{ij} (\vec{k}, \omega) = 
        2D_{th}\sum_{\nu = \{x,y\}}S_{i\nu}(\vec{k}, \omega)S_{j\nu}(-\vec{k}, -\omega)
            \label{sup_eq:correlations}
\end{align}
where $D_\text{th} = \frac{\alpha k_B T}{\gamma M_\text{s} L}$, with $k_B$ the Boltzmann constant, $T$ the temperature, $S$ the magnetic susceptibility (Eq. \ref{sup_eq:susceptibility_matrix}).

The dipolar tensor $D^\text{eff}(\vec{k},\omega)$ is obtained by first rotating the magnet frame to the lab frame, then multiplying by the dipolar tensor $\mathcal{D}(\vec{k})$ in the lab frame, and then rotating the result to the NV frame: $D^\text{eff}(\vec{k},\omega) = R_{yz}(\theta_\text{NV}, \phi_\text{NV}) \mathcal{D}(\vec{k})R_Y(\theta_0)^T$, where 
\begin{align}
    \mathcal{D}(\vec{k})
        =
        -\frac{\mu_0 M_s}{2} e^{-|\vec{k}| d_\text{NV}} (1 - e^{-|\vec{k}|L})
        \begin{bmatrix}
            \cos^2 \phi_k & \sin(2\phi_k)/2 & i\cos \phi_k \\ 
            \sin(2\phi_k)/2 & \sin^2 \phi_k & i\sin \phi_k \\
            i\cos \phi_k & i\sin \phi_k & -1
        \end{bmatrix},
    \label{sup_eq:dipolar_tensor_in_lab_frame}
\end{align}
where $\mu_0$ is the vacuum permeability and $d_\text{NV}$ is the distance between the NV and the sample surface. The terms in Eq. \eqref{eq:NV_relaxation_rate} that induce spin relaxation are given by\cite{Rustagi2020}: $\mathcal{D}^\text{eff}_{\pm \nu} = \mathcal{D}^\text{eff}_{x \nu} \pm i\mathcal{D}^\text{eff}_{y \nu}$.

For a magnon gas in thermal equilibrium, in the absence of microwave driving, the dependence of the NV relaxation rate on the NV-sample distance can be calculated using Eq. \eqref{eq:NV_relaxation_rate}. The fast increase in rate (Fig. \ref{Fig_2}d, main text) results in a reduction of the  ESR contrast as the NV sensor approaches the film to within nanometer proximity.

\subsection{Spatial ESR contrast generated by a single spin wave}\label{sup_sec:spatial_Cesr}
\begin{figure}[H]
    \centering
    \includegraphics{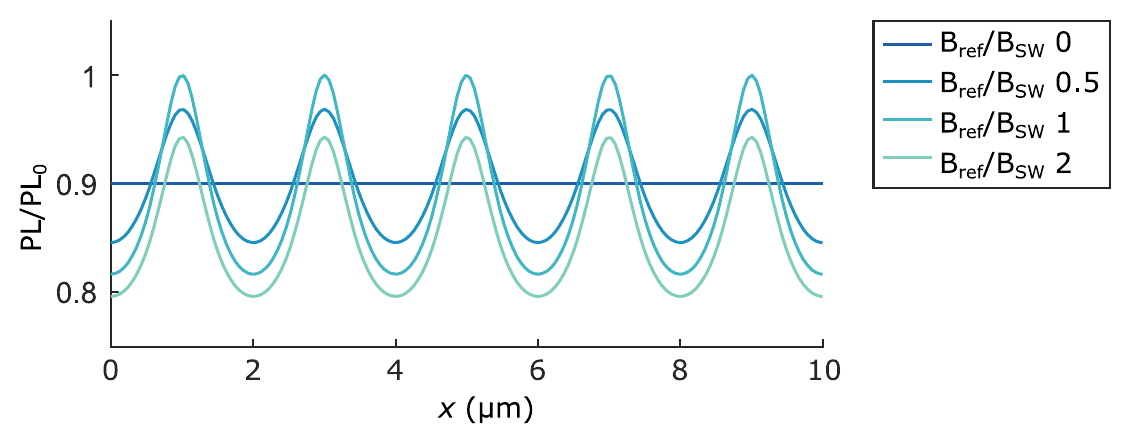}
    \caption{\textbf{Spatial profile of a single spin wave mode interfering with a reference field.} The expected $\mathrm{PL}/\mathrm{PL}_0$ for a single spin-wave mode for various strengths of the reference field.}
    \label{sup_fig:Bref_Bsw}
\end{figure}

Here, we determine the spatial profile of the ESR contrast generated by a propagating spin-wave mode, with wavenumber $k$ that interferes with an uniform reference field of varying amplitude (Fig. \ref{sup_fig:Bref_Bsw})\cite{Zhou2021, Bertelli2020}. We show that the ESR contrast spatially varies depending on the wavenumber given that the reference field has a finite amplitude.

Spin waves produce a field of which the component ($B_{\text{SW}}$) that is rotating with the correct handedness in a plane that is perpendicular to the NV-axis drives Rabi oscillations\cite{Bertelli2020}.
This component varies spatially according to:
\begin{equation}
    B_{\text{SW}} = B^0_{\text{SW}}e^{ik(x-x_0)}
\end{equation}

This field induces NV Rabi oscillations of which the rate is given by: 
\begin{equation}
    \Omega_R = \frac{\gamma}{\sqrt{2}}|B_{\text{SW}} + B_{\text{ref}}|
\label{sup_eq:Rabifreq}
\end{equation}
where $B_{\text{ref}}$ is the component of the reference field that is rotating with the correct handedness in a plane perpendicular to the NV-axis. The ESR contrast is given by:
\begin{align}
    \mathrm{PL}/\mathrm{PL}_0 &= 1 - C_{\text{ESR}}\\
    \mathrm{PL}/\mathrm{PL}_0 &= 1 - \beta \frac{\Omega^2_R}{\Omega^2_R + \delta^2}
    \label{sup_eq:ESR_contrast}
\end{align}
where $\beta$ is a constant that describes the maximum ESR contrast and $\delta$ is a parameter that depends on the optical pumping rate\cite{Dreau2011}, which we assume to be constant as all measurements were taken at the same laser power.

\subsubsection{Extracting the spin wavelength of the low wavenumber mode}
\begin{figure}[H]
    \centering
    \includegraphics{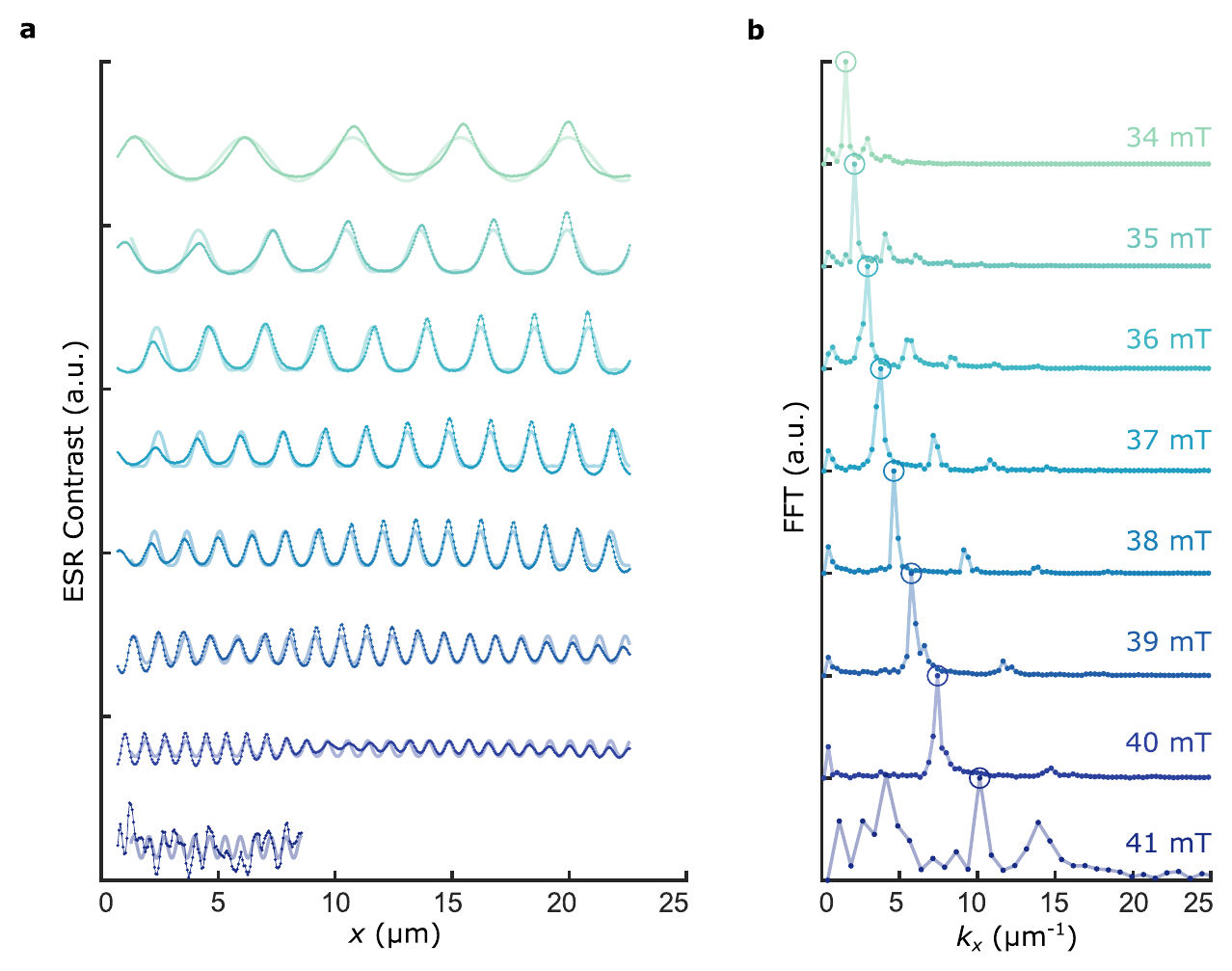}
    \caption{\textbf{Wavelength analysis of spin-wave maps} (a) We average the 2D maps of figure 3b of the main text, to obtain a 1D linetrace of the ESR contrast as a function of $x$-position. We fit the data using equation \ref{sup_eq:ESR_contrast}, to obtain the wavenumber and its uncertainty. (b) Fourier transform of the 1D data shown in (a). Circles represent peak positions used as initial guess for the fitting.}
    \label{sup_fig:wavelength_extraction}
\end{figure}

Here we analyze the spatial maps shown in figure 3b of the main text and extract the wavelength of the imaged modes. 
To do so, we first plot the signal as a function of $x$ (Fig. \ref{sup_fig:wavelength_extraction}a), after subtracting a linear term to account for the non-uniformity of the stripline-field. Using equation \ref{sup_eq:ESR_contrast}, we then fit the averaged data (Fig. \ref{sup_fig:wavelength_extraction}a). The fit allows to extract the wavenumber $k$, which we plot as a function of $B_0$ (Fig. \ref{Fig_3}c, main text). In figure \ref{sup_fig:wavelength_extraction}b, we show the corresponding Fourier transform of the averaged data traces.
Finally, we fit the extracted wavenumbers to the backward-volume spin-wave dispersion and we find that an angle of $\theta_B = 49$\textdegree{}, which is the angle between the magnetic field and the YIG surface normal, fits our data best, due to an uncertainty in mounting of the NV-probe with respect to the sample surface.

\subsection{Combined atomic force microscopy and photoluminescence scans of the YIG surface}\label{sup_sec:yig_surface}
\begin{figure}[H]
    \centering
    \includegraphics{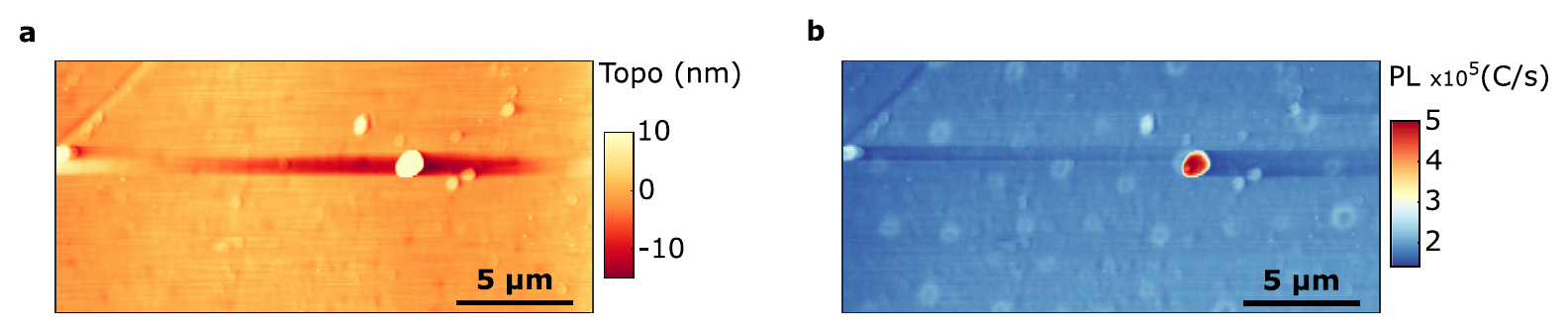}
    \caption{\textbf{YIG surface.} (a) Surface topography (Topo) and (b) photoluminescence (PL) of our 235 nm-thick YIG film surface. Data corresponds to the ESR map in figure \ref{sup_fig:overview_spatialscans}b, scan 1 taken with the tip in contact with the YIG surface.}
    \label{sup_fig:yig_surface}
\end{figure}

In contact mode, our scanning NV magnetometry setup collects both the topography, via the AFM feedback signal, and the NV photoluminescence  (Fig. \ref{sup_fig:yig_surface}). The topography image not only shows dirt particles laying on top of the surface (large white dot in the center of the scan), but  also scratches and tiny pits in the YIG surface that can lead to magnon scattering \cite{Gross2020, Graefe2020}.

\subsection{Overview spin-wave images}\label{sup_sec:overview}
\begin{figure}[H]
    \centering
    \includegraphics[scale=0.97]{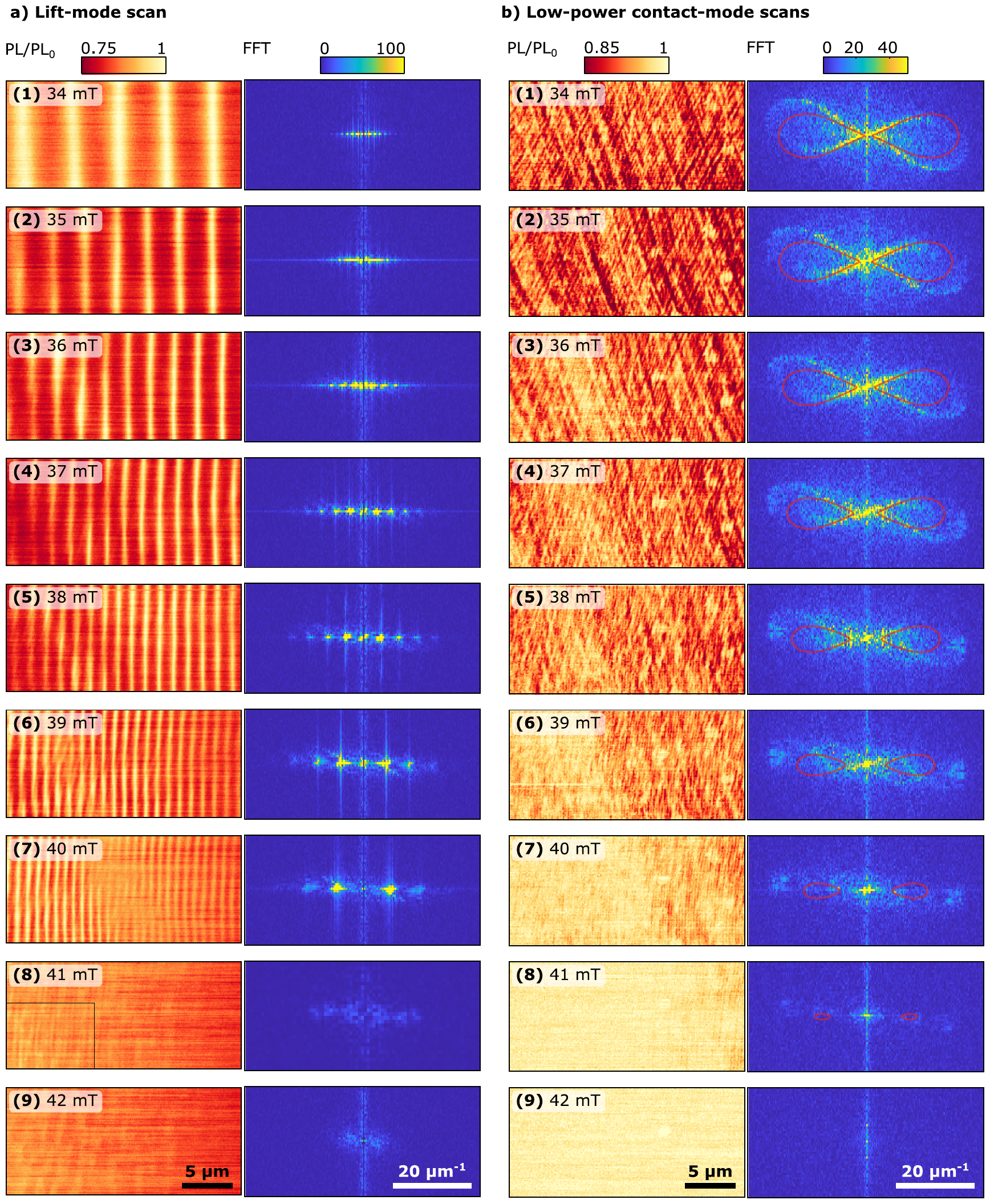}
    \caption{\textbf{Overview spin-wave images} (a) Scans corresponding to Fig. \ref{Fig_3}b of the main text using $P_{\mathrm{MW}}= \SI{4}{\milli\watt}$ and we change the NV-to-sample depending on the spin-wavelength. (b) Scans corresponding to Fig. \ref{Fig_4}a of the main text using $P_{\mathrm{MW}}= \SI{6.3}{\micro\watt}$ while keeping the NV-tip in contact with the magnetic surface. Red line: the calculated isofrequency contour of the 2D spin-wave dispersion.} 
    \label{sup_fig:overview_spatialscans}
\end{figure}

\subsection{Calibration of the piezoelectric scanners}\label{sup_sec:calibration}
\subsubsection{Lateral displacement}
\begin{figure}[H]
    \centering
    \includegraphics{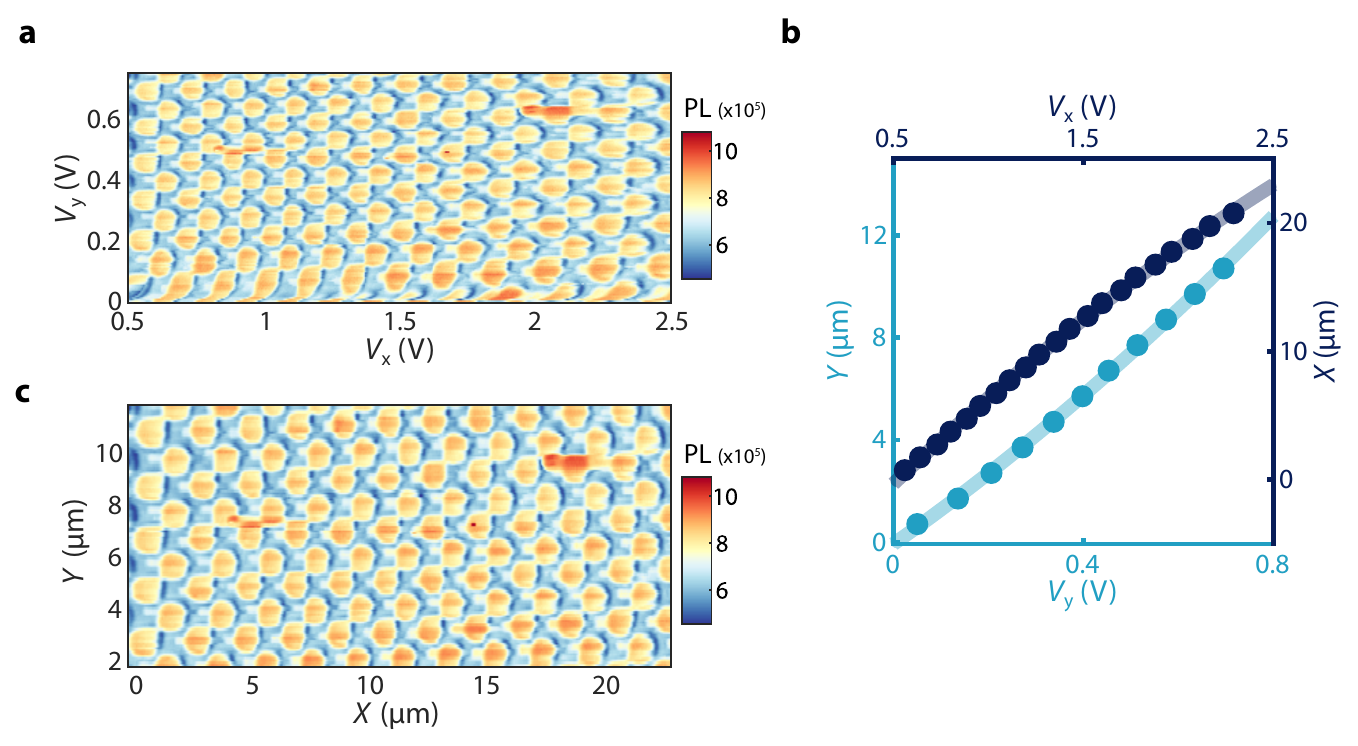}
    \caption{\textbf{Calibration of lateral piezoelectric scanners.} (a) 2D spatial scan of calibration sample with a \SI{2}{\um} pitch using linearly spaced voltage intervals at a scan speed of \SI{0.1}{\V\per\s}. (b) We fit the known X and Y-displacement of the chess pits (using a single row or column) as a function of applied X and Y voltage. (c) Interpolated data shown in (a), now using linearly spaced position intervals obtained by the fitting functions in (b) and with the first 25 scan lines removed}
    \label{sup_fig:scanner_calib}
\end{figure}

 Our piezoelectric scanner (ANSxy50/LT) exhibits a nonlinear motion as a function of applied voltage. We calibrate this non-linear motion using a silicon nitride sample with \SI{2}{\um} chess pitch structures. Specifically, we scan over the same region of interest and piezoscanner voltages/offsets while recording the photoluminescence (Figure \ref{sup_fig:scanner_calib}a). Our calibration procedure to convert the non-linear displacement as a function of applied voltage to position is as follows:
\begin{enumerate}
    \item We first remove the first 25 lines from our scan data which show large non-linear and non-reproducible displacements depending on scan-speed and time spent on the first pixel. 
    \item We assign the applied voltage to known positions of subsequent chess pitches (Fig. \ref{sup_fig:scanner_calib}). For simplicity, we do this for a single row or column (Fig. \ref{sup_fig:scanner_calib}b).
    \item We fit a second order polynomial to the known X or Y displacement. 
\end{enumerate}
We repeat this process for the two scan speeds used in this work: \SI{0.1}{\V\per\s} and \SI{0.05}{\V\per\s} used for the lift-mode and contact mode scans. We can now interpolate our 2D spatial scans, that were taken at linearly spaced voltage interval and obtain a 2D image with linearly spaced position intervals. Note that all scan data shown in the manuscript are taken in the forward scan direction and we therefore do not take piezoelectric hysteresis into account. 

\subsubsection{Calibration of the NV-to-sample distance}
\begin{figure}[H]
    \centering
    \includegraphics{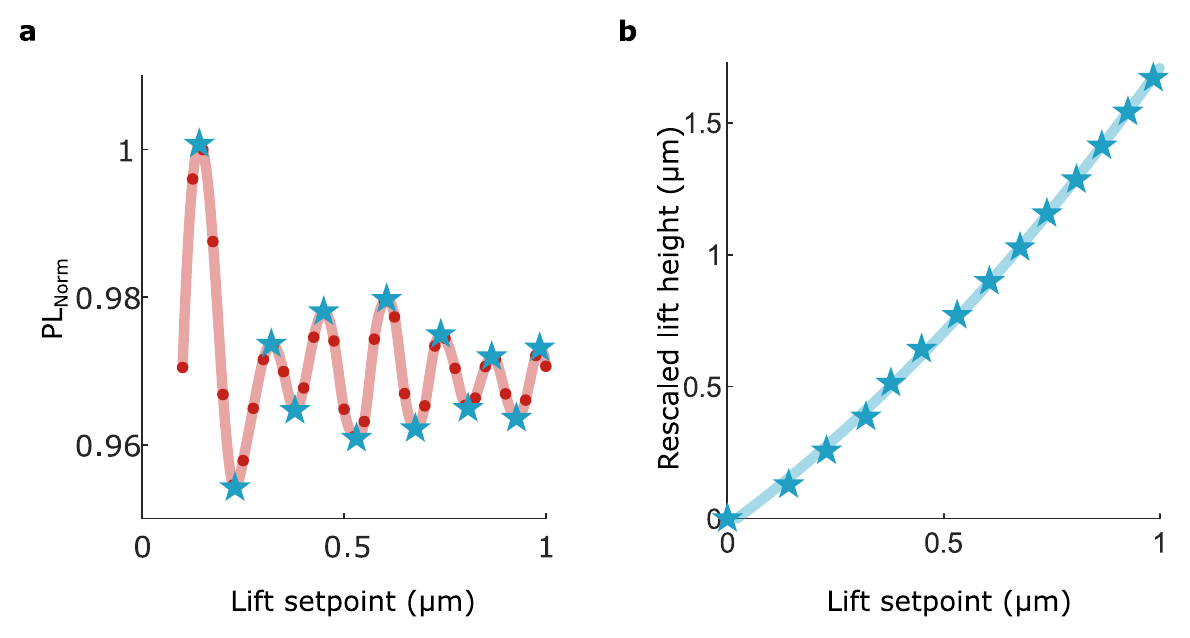}
    \caption{\textbf{Calibration of lift height.} (a) Oscillating PL signal when the tip is retracted from the YIG surface. The stars are the found extrema. Data corresponds to the Rabi measurement shown in Fig. \ref{Fig_2}b of the main text. (b) We allocate a distance of $\lambda/4$ between each extremum (stars), of which the first maximum corresponds to an absolute distance of $\lambda/4$ from the surface. Our contact position is fixed at zero lift height. We use a second order polynomial to fit the displacement (solid line).}
    \label{sup_fig:calibration_z_scanner}
\end{figure}

When retracting the NV-tip from the YIG surface, we observe oscillations in the NV PL (Fig. \ref{sup_fig:calibration_z_scanner}a). We assume that the oscillations are caused by interference between the YIG-surface-reflected laser light and the laser light internally reflected in the diamond tip (akin to the effect leading to Newton rings). The interference leads to PL oscillations with a spatial period equal to half the laser wavelength (i.e. \SI{515}{\nano\m}/2). We use these oscillations to estimate the tip-sample distance $d$ for each setpoint distance obtained from the linear voltage-to-distance conversion provided by the supplier of the scanners (Fig. \ref{sup_fig:calibration_z_scanner}b) .
\printbibliography

@Article{Acremann2000,
  author  = {Acremann, Y. and Back, C. H. and Buess, M. and Portmann, O. and Vaterlaus, A. and Pescia, D. and Melchior, H.},
  journal = {Science},
  title   = {Imaging Precessional Motion of the Magnetization Vector},
  year    = {2000},
  number  = {5491},
  pages   = {492-495},
  volume  = {290},
  doi     = {doi:10.1126/science.290.5491.492},
  type    = {Journal Article},
  url     = {https://www.science.org/doi/abs/10.1126/science.290.5491.492 %X We report on imaging of three-dimensional precessional orbits of the magnetization vector in a magnetic field by means of a time-resolved vectorial Kerr experiment that measures all three components of the magnetization vector with picosecond resolution. Images of the precessional mode taken with submicrometer spatial resolution reveal that the dynamical excitation in this time regime roughly mirrors the symmetry of the underlying equilibrium spin configuration and that its propagation has a non-wavelike character. These results should form the basis for realistic models of the magnetization dynamics in a largely unexplored but technologically increasingly relevant time scale.},
}

@Article{Andrich2017,
  author  = {Andrich, Paolo and de las Casas, Charles F. and Liu, Xiaoying and Bretscher, Hope L. and Berman, Jonson R. and Heremans, F. Joseph and Nealey, Paul F. and Awschalom, David D.},
  journal = {npj Quantum Information},
  title   = {Long-range spin wave mediated control of defect qubits in nanodiamonds},
  year    = {2017},
  issn    = {2056-6387},
  pages   = {28},
  volume  = {3},
  doi     = {10.1038/s41534-017-0029-z},
  type    = {Journal Article},
  url     = {https://doi.org/10.1038/s41534-017-0029-z},
}

@Article{Bertelli2020,
  author  = {Bertelli, Iacopo and Carmiggelt, Joris J and Yu, Tao and Simon, Brecht G and Pothoven, Coosje C and Bauer, Gerrit EW and Blanter, Yaroslav M and Aarts, Jan and Van Der Sar, Toeno},
  journal = {Science Advances},
  title   = {Magnetic resonance imaging of spin-wave transport and interference in a magnetic insulator},
  year    = {2020},
  issn    = {2375-2548},
  number  = {46},
  pages   = {eabd3556},
  volume  = {6},
  type    = {Journal Article},
}

@Article{Bertelli2021a,
  author  = {Bertelli, Iacopo and Simon, Brecht G. and Yu, Tao and Aarts, Jan and Bauer, Gerrit E. W. and Blanter, Yaroslav M. and van der Sar, Toeno},
  journal = {Advanced Quantum Technologies},
  title   = {Imaging Spin-Wave Damping Underneath Metals Using Electron Spins in Diamond},
  year    = {2021},
  number  = {12},
  pages   = {2100094},
  volume  = {4},
  doi     = {https://doi.org/10.1002/qute.202100094},
  type    = {Journal Article},
  url     = {https://onlinelibrary.wiley.com/doi/abs/10.1002/qute.202100094},
}

@Article{Chumak2022,
  author  = {Chumak, A. V. and Kabos, P. and Wu, M. and Abert, C. and Adelmann, C. and Adeyeye, A. and Åkerman, J. and Aliev, F. G. and Anane, A. and Awad, A. and Back, C. H. and Barman, A. and Bauer, G. E. W. and Becherer, M. and Beginin, E. N. and Bittencourt, V. A. S. V. and Blanter, Y. M. and Bortolotti, P. and Boventer, I. and Bozhko, D. A. and Bunyaev, S. A. and Carmiggelt, J. J. and Cheenikundil, R. R. and Ciubotaru, F. and Cotofana, S. and Csaba, G. and Dobrovolskiy, O. V. and Dubs, C. and Elyasi, M. and Fripp, K. G. and Fulara, H. and Golovchanskiy, I. A. and Gonzalez-Ballestero, C. and Graczyk, P. and Grundler, D. and Gruszecki, P. and Gubbiotti, G. and Guslienko, K. and Haldar, A. and Hamdioui, S. and Hertel, R. and Hillebrands, B. and Hioki, T. and Houshang, A. and Hu, C. M. and Huebl, H. and Huth, M. and Iacocca, E. and Jungfleisch, M. B. and Kakazei, G. N. and Khitun, A. and Khymyn, R. and Kikkawa, T. and Kläui, M. and Klein, O. and Kłos, J. W. and Knauer, S. and Koraltan, S. and Kostylev, M. and Krawczyk, M. and Krivorotov, I. N. and Kruglyak, V. V. and Lachance-Quirion, D. and Ladak, S. and Lebrun, R. and Li, Y. and Lindner, M. and Macêdo, R. and Mayr, S. and Melkov, G. A. and Mieszczak, S. and Nakamura, Y. and Nembach, H. T. and Nikitin, A. A. and Nikitov, S. A. and Novosad, V. and Otálora, J. A. and Otani, Y. and Papp, A. and Pigeau, B. and Pirro, P. and Porod, W. and Porrati, F. and Qin, H. and Rana, B. and Reimann, T. and Riente, F. and Romero-Isart, O. and Ross, A. and Sadovnikov, A. V. and Safin, A. R. and Saitoh, E. and Schmidt, G. and Schultheiss, H. and Schultheiss, K. and Serga, A. A. and Sharma, S. and Shaw, J. M. and Suess, D. and Surzhenko, O. and others},
  journal = {IEEE Transactions on Magnetics},
  title   = {Roadmap on Spin-Wave Computing},
  year    = {2022},
  issn    = {1941-0069},
  number  = {6},
  pages   = {0800172},
  volume  = {58},
  doi     = {10.1109/TMAG.2022.3149664},
  type    = {Journal Article},
}

@Article{Chumak2014,
  author  = {Chumak, Andrii V and Serga, Alexander A and Hillebrands, Burkard},
  journal = {Nature Communications},
  title   = {Magnon transistor for all-magnon data processing},
  year    = {2014},
  issn    = {2041-1723},
  pages   = {4700},
  volume  = {5},
  type    = {Journal Article},
}

@Article{Cornelissen2015,
  author  = {Cornelissen, LJ and Liu, J and Duine, RA and Youssef, J Ben and Van Wees, BJ},
  journal = {Nature Physics},
  title   = {Long-distance transport of magnon spin information in a magnetic insulator at room temperature},
  year    = {2015},
  issn    = {1745-2481},
  pages   = {1022-1026},
  volume  = {11},
  type    = {Journal Article},
}

@Article{Degen2017,
  author  = {Degen, C.  L and Reinhard, F. and Cappellaro, P.},
  journal = {Reviews of Modern Physics},
  title   = {Quantum sensing},
  year    = {2017},
  number  = {3},
  pages   = {035002},
  volume  = {89},
  doi     = {10.1103/RevModPhys.89.035002},
  type    = {Journal Article},
  url     = {https://link.aps.org/doi/10.1103/RevModPhys.89.035002},
}

@Article{Demokritov2006,
  author  = {Demokritov, Sergej O and Demidov, Vladislav E and Dzyapko, Oleksandr and Melkov, Gennadii A and Serga, Alexandar A and Hillebrands, Burkard and Slavin, Andrei N},
  journal = {Nature},
  title   = {Bose–Einstein condensation of quasi-equilibrium magnons at room temperature under pumping},
  year    = {2006},
  issn    = {1476-4687},
  pages   = {430-433},
  volume  = {443},
  type    = {Journal Article},
}

@Article{Dreau2011,
  author  = {Dréau, A. and Lesik, M. and Rondin, L. and Spinicelli, P. and Arcizet, O. and Roch, J. F. and Jacques, V.},
  journal = {Physical Review B},
  title   = {Avoiding power broadening in optically detected magnetic resonance of single NV defects for enhanced dc magnetic field sensitivity},
  year    = {2011},
  number  = {19},
  pages   = {195204},
  volume  = {84},
  doi     = {10.1103/PhysRevB.84.195204},
  type    = {Journal Article},
  url     = {https://link.aps.org/doi/10.1103/PhysRevB.84.195204},
}

@Article{Du2017,
  author  = {Du, Chunhui and Van der Sar, Toeno and Zhou, Tony X and Upadhyaya, Pramey and Casola, Francesco and Zhang, Huiliang and Onbasli, Mehmet C and Ross, Caroline A and Walsworth, Ronald L and Tserkovnyak, Yaroslav},
  journal = {Science},
  title   = {Control and local measurement of the spin chemical potential in a magnetic insulator},
  year    = {2017},
  issn    = {0036-8075},
  number  = {6347},
  pages   = {195-198},
  volume  = {357},
  type    = {Journal Article},
}

@Article{Finco2021,
  author  = {Finco, Aurore and Haykal, Angela and Tanos, Rana and Fabre, Florentin and Chouaieb, Saddem and Akhtar, Waseem and Robert-Philip, Isabelle and Legrand, William and Ajejas, Fernando and Bouzehouane, Karim},
  journal = {Nature communications},
  title   = {Imaging non-collinear antiferromagnetic textures via single spin relaxometry},
  year    = {2021},
  issn    = {2041-1723},
  pages   = {767},
  volume  = {12},
  type    = {Journal Article},
}

@Article{Flebus2018,
  author  = {Flebus, B. and Tserkovnyak, Y.},
  journal = {Physical Review Letters},
  title   = {Quantum-Impurity Relaxometry of Magnetization Dynamics},
  year    = {2018},
  number  = {18},
  pages   = {187204},
  volume  = {121},
  doi     = {10.1103/PhysRevLett.121.187204},
  type    = {Journal Article},
  url     = {https://link.aps.org/doi/10.1103/PhysRevLett.121.187204},
}

@Article{Graefe2020,
  author  = {Gräfe, Joachim and Gruszecki, Pawel and Zelent, Mateusz and Decker, Martin and Keskinbora, Kahraman and Noske, Matthias and Gawronski, Przemysław and Stoll, Hermann and Weigand, Markus and Krawczyk, Maciej and Back, Christian H. and Goering, Eberhard J. and Schütz, Gisela},
  journal = {Physical Review B},
  title   = {Direct observation of spin-wave focusing by a Fresnel lens},
  year    = {2020},
  number  = {2},
  pages   = {024420},
  volume  = {102},
  doi     = {10.1103/PhysRevB.102.024420},
  type    = {Journal Article},
  url     = {https://link.aps.org/doi/10.1103/PhysRevB.102.024420},
}

@Article{Gross2020,
  author  = {Groß, Felix and Zelent, Mateusz and Träger, Nick and Förster, Johannes and Sanli, Umut T. and Sauter, Robert and Decker, Martin and Back, Christian H. and Weigand, Markus and Keskinbora, Kahraman and Schütz, Gisela and Krawczyk, Maciej and Gräfe, Joachim},
  journal = {ACS Nano},
  title   = {Building Blocks for Magnon Optics: Emission and Conversion of Short Spin Waves},
  year    = {2020},
  issn    = {1936-0851},
  number  = {12},
  pages   = {17184-17193},
  volume  = {14},
  doi     = {10.1021/acsnano.0c07076},
  type    = {Journal Article},
  url     = {https://doi.org/10.1021/acsnano.0c07076},
}

@Article{Hula2020,
  author  = {Hula, Tobias and Schultheiss, Katrin and Buzdakov, Aleksandr and Körber, Lukas and Bejarano, Mauricio and Flacke, Luis and Liensberger, Lukas and Weiler, Mathias and Shaw, Justin M. and Nembach, Hans T. and Fassbender, Jürgen and Schultheiss, Helmut},
  journal = {Applied Physics Letters},
  title   = {Nonlinear losses in magnon transport due to four-magnon scattering},
  year    = {2020},
  number  = {4},
  pages   = {042404},
  volume  = {117},
  doi     = {10.1063/5.0015269},
  type    = {Journal Article},
  url     = {https://aip.scitation.org/doi/abs/10.1063/5.0015269},
}

@Article{Koerner2022,
  author  = {Koerner, Chris and Dreyer, Rouven and Wagener, Martin and Liebing, Niklas and Bauer, Hans G. and Woltersdorf, Georg},
  journal = {Science},
  title   = {Frequency multiplication by collective nanoscale spin-wave dynamics},
  year    = {2022},
  number  = {6585},
  pages   = {1165-1169},
  volume  = {375},
  doi     = {doi:10.1126/science.abm6044},
  type    = {Journal Article},
  url     = {https://www.science.org/doi/abs/10.1126/science.abm6044 %X Frequency multiplication is a process in modern electronics in which harmonics of the input frequency are generated in nonlinear electronic circuits. Devices based on the propagation and interaction of spin waves are a promising alternative to conventional electronics. The characteristic frequency of these excitations is in the gigahertz (GHz) range and devices are not readily interfaced with conventional electronics. Here, we locally probe the magnetic excitations in a soft magnetic material by optical methods and show that megahertz-range excitation frequencies cause switching effects on the micrometer scale, leading to phase-locked spin-wave emission in the GHz range. Indeed, the frequency multiplication process inside the magnetic medium covers six octaves and opens exciting perspectives for spintronic applications, such as all-magnetic mixers or on-chip GHz sources. The generation and propagation of magnetic excitations such as magnons and spin waves in ferromagnetic thin films provides a platform for the development of spin-based device technology. Koerner et al. report measurements on the magnetization dynamics of a nickel–iron film excited coherently by microwave magnetic fields from a coplanar wave guide. Using nitrogen vacancy center–based magnetometry and a time-resolved magneto-optical Kerr effect, the authors show that low excitation frequencies and low bias fields in the range of only a few milli-tesla results in the generation of magnons emitted at higher frequency. Extending over 60 harmonics of the excitation frequency, such upconversion of magnetic excitation frequencies should prove useful for spintronics applications. —ISO Magnetic excitations in a thin ferromagnetic film can be upconverted from megahertz to gigahertz frequencies.},
}

@Article{McCullian2020,
  author  = {McCullian, Brendan A and Thabt, Ahmed M and Gray, Benjamin A and Melendez, Alex L and Wolf, Michael S and Safonov, Vladimir L and Pelekhov, Denis V and Bhallamudi, Vidya P and Page, Michael R and Hammel, P Chris},
  journal = {Nature Communications},
  title   = {Broadband multi-magnon relaxometry using a quantum spin sensor for high frequency ferromagnetic dynamics sensing},
  year    = {2020},
  issn    = {2041-1723},
  pages   = {5229},
  volume  = {11},
  type    = {Journal Article},
}

@Article{Mohseni2019,
  author  = {Mohseni, M. and Verba, R. and Brächer, T. and Wang, Q. and Bozhko, D.  A and Hillebrands, B. and Pirro, P.},
  journal = {Physical Review Letters},
  title   = {Backscattering Immunity of Dipole-Exchange Magnetostatic Surface Spin Waves},
  year    = {2019},
  number  = {19},
  pages   = {197201},
  volume  = {122},
  doi     = {10.1103/PhysRevLett.122.197201},
  type    = {Journal Article},
  url     = {https://link.aps.org/doi/10.1103/PhysRevLett.122.197201},
}

@Book{Rezende2020,
  author    = {Rezende, Sergio M},
  publisher = {Springer},
  title     = {Fundamentals of magnonics},
  year      = {2020},
  volume    = {969},
  type      = {Book},
}

@Article{Rondin2014,
  author  = {Rondin, L. and Tetienne, J. P. and Hingant, T. and Roch, J. F. and Maletinsky, P. and Jacques, V.},
  journal = {Reports on Progress in Physics},
  title   = {Magnetometry with nitrogen-vacancy defects in diamond},
  year    = {2014},
  issn    = {0034-4885
1361-6633},
  number  = {5},
  pages   = {056503},
  volume  = {77},
  doi     = {10.1088/0034-4885/77/5/056503},
  type    = {Journal Article},
  url     = {http://dx.doi.org/10.1088/0034-4885/77/5/056503},
}

@Article{Rustagi2020,
  author  = {Rustagi, Avinash and Bertelli, Iacopo and Van Der Sar, Toeno and Upadhyaya, Pramey},
  journal = {Physical Review B},
  title   = {Sensing chiral magnetic noise via quantum impurity relaxometry},
  year    = {2020},
  number  = {22},
  pages   = {220403(R)},
  volume  = {102},
  type    = {Journal Article},
}

@Article{Sebastian2015,
  author  = {Sebastian, Thomas and Schultheiss, Katrin and Obry, Björn and Hillebrands, Burkard and Schultheiss, Helmut},
  journal = {Frontiers in Physics},
  title   = {Micro-focused Brillouin light scattering: imaging spin waves at the nanoscale},
  year    = {2015},
  issn    = {2296-424X},
  pages   = {35},
  volume  = {3},
  type    = {Journal Article},
}

@Article{Serga2010,
  author  = {Serga, AA and Chumak, AV and Hillebrands, B},
  journal = {Journal of Physics D: Applied Physics},
  title   = {YIG magnonics},
  year    = {2010},
  issn    = {0022-3727},
  number  = {26},
  pages   = {264002},
  volume  = {43},
  type    = {Journal Article},
}

@Article{Sluka2019,
  author  = {Sluka, Volker and Schneider, Tobias and Gallardo, Rodolfo A. and Kákay, Attila and Weigand, Markus and Warnatz, Tobias and Mattheis, Roland and Roldán-Molina, Alejandro and Landeros, Pedro and Tiberkevich, Vasil and Slavin, Andrei and Schütz, Gisela and Erbe, Artur and Deac, Alina and Lindner, Jürgen and Raabe, Jörg and Fassbender, Jürgen and Wintz, Sebastian},
  journal = {Nature Nanotechnology},
  title   = {Emission and propagation of 1D and 2D spin waves with nanoscale wavelengths in anisotropic spin textures},
  year    = {2019},
  issn    = {1748-3395},
  pages   = {328-333},
  volume  = {14},
  doi     = {10.1038/s41565-019-0383-4},
  type    = {Journal Article},
  url     = {https://doi.org/10.1038/s41565-019-0383-4},
}

@Article{Wang2018,
  author  = {Wang, Qi and Pirro, Philipp and Verba, Roman and Slavin, Andrei and Hillebrands, Burkard and Chumak, Andrii V.},
  journal = {Science Advances},
  title   = {Reconfigurable nanoscale spin-wave directional coupler},
  year    = {2018},
  number  = {1},
  pages   = {e1701517},
  volume  = {4},
  doi     = {doi:10.1126/sciadv.1701517},
  type    = {Journal Article},
  url     = {https://www.science.org/doi/abs/10.1126/sciadv.1701517 %X We propose a nanoscale spin-wave directional coupler that allows the realization of magnonic integrated circuits. Spin waves, and their quanta magnons, are prospective data carriers in future signal processing systems because Gilbert damping associated with the spin-wave propagation can be made substantially lower than the Joule heat losses in electronic devices. Although individual spin-wave signal processing devices have been successfully developed, the challenging contemporary problem is the formation of two-dimensional planar integrated spin-wave circuits. Using both micromagnetic modeling and analytical theory, we present an effective solution of this problem based on the dipolar interaction between two laterally adjacent nanoscale spin-wave waveguides. The developed device based on this principle can work as a multifunctional and dynamically reconfigurable signal directional coupler performing the functions of a waveguide crossing element, tunable power splitter, frequency separator, or multiplexer. The proposed design of a spin-wave directional coupler can be used both in digital logic circuits intended for spin-wave computing and in analog microwave signal processing devices.},
}

@Article{Wolfe2014,
  author  = {Wolfe, Chris S and Bhallamudi, Vidya P and Wang, Hilong L and Du, Chunhui H and Manuilov, Sergei and Teeling-Smith, RM and Berger, AJ and Adur, R and Yang, FY and Hammel, Peter Christopher},
  journal = {Physical Review B},
  title   = {Off-resonant manipulation of spins in diamond via precessing magnetization of a proximal ferromagnet},
  year    = {2014},
  number  = {18},
  pages   = {180406(R)},
  volume  = {89},
  type    = {Journal Article},
}

@Article{Zhou2021,
  author  = {Zhou, Tony X. and Carmiggelt, Joris J. and Gächter, Lisa M. and Esterlis, Ilya and Sels, Dries and Stöhr, Rainer J. and Du, Chunhui and Fernandez, Daniel and Rodriguez-Nieva, Joaquin F. and Büttner, Felix and Demler, Eugene and Yacoby, Amir},
  journal = {Proceedings of the National Academy of Sciences},
  title   = {A magnon scattering platform},
  year    = {2021},
  number  = {25},
  pages   = {e2019473118},
  volume  = {118},
  doi     = {doi:10.1073/pnas.2019473118},
  type    = {Journal Article},
  url     = {https://www.pnas.org/doi/abs/10.1073/pnas.2019473118 %X This work describes a general scattering platform that uses magnons to explore the underlying properties of target materials. In this work we show how both phase and amplitude of magnons can be imaged using a nitrogen vacancy center magnetometer and how the scattered pattern of waves can be used to infer geometric and magnetic properties of a target material. To demonstrate this new experimental methodology we use a permalloy disk as our target and show that even with such a simple target unexpected behavior is observed. In addition, we provide a theoretical framework to reconstruct the properties of the target. Scattering experiments have revolutionized our understanding of nature. Examples include the discovery of the nucleus [R. G. Newton, Scattering Theory of Waves and Particles (1982)], crystallography [U. Pietsch, V. Holý, T. Baumback, High-Resolution X-Ray Scattering (2004)], and the discovery of the double-helix structure of DNA [J. D. Watson, F. H. C. Crick, Nature 171, 737–738]. Scattering techniques differ by the type of particles used, the interaction these particles have with target materials, and the range of wavelengths used. Here, we demonstrate a two-dimensional table-top scattering platform for exploring magnetic properties of materials on mesoscopic length scales. Long-lived, coherent magnonic excitations are generated in a thin film of yttrium iron garnet and scattered off a magnetic target deposited on its surface. The scattered waves are then recorded using a scanning nitrogen vacancy center magnetometer that allows subwavelength imaging and operation under conditions ranging from cryogenic to ambient environment. While most scattering platforms measure only the intensity of the scattered waves, our imaging method allows for spatial determination of both amplitude and phase of the scattered waves, thereby allowing for a systematic reconstruction of the target scattering potential. Our experimental results are consistent with theoretical predictions for such a geometry and reveal several unusual features of the magnetic response of the target, including suppression near the target edges and a gradient in the direction perpendicular to the direction of surface wave propagation. Our results establish magnon scattering experiments as a platform for studying correlated many-body systems.},
}
\end{refsection} 
\end{document}